\documentclass[%
 reprint,
 superscriptaddress,
 bibnotes,
 amsmath,amssymb,
 aps,
 prl,
 floatfix]{revtex4-2}

\usepackage[dvipdfmx]{graphicx}
\usepackage{bm}
\usepackage{hyperref}
\usepackage{float}
\hypersetup{
  colorlinks   = true,
  urlcolor     = black,
  linkcolor    = black,
  citecolor   = blue
}
\usepackage{etoolbox}
\begin{document}

\title{Three-dimensional zigzag correlations in the van der Waals Kitaev magnet RuBr$_3$}

\author{H. Gretarsson}
\email[]{hlynur.gretarsson@desy.de}
\affiliation{Deutsches Elektronen-Synchrotron DESY, Notkestrstra\ss e 85, D-22607 Hamburg, Germany}
\author{R. Iwazaki}
\affiliation{Department of Physics, Graduate School of Science, Tohoku University, 6-3 Aramaki-Aoba, Aoba-ku, Sendai, Miyagi 980-8578, Japan}
\author{F. Sato}
\affiliation{Department of Physics, Graduate School of Science, Tohoku University, 6-3 Aramaki-Aoba, Aoba-ku, Sendai, Miyagi 980-8578, Japan}
\author{H. Gotou}
\affiliation{Institute for Solid State Physics (ISSP), University of Tokyo, Kashiwa, Chiba 277-8581, Japan}
\author{S. Francoual}
\affiliation{Deutsches Elektronen-Synchrotron DESY, Notkestrstra\ss e 85, D-22607 Hamburg, Germany}
\author{J. Nasu}
\affiliation{Department of Physics, Graduate School of Science, Tohoku University, 6-3 Aramaki-Aoba, Aoba-ku, Sendai, Miyagi 980-8578, Japan}
\author{Y. Imai}
\affiliation{Department of Physics, Graduate School of Science, Tohoku University, 6-3 Aramaki-Aoba, Aoba-ku, Sendai, Miyagi 980-8578, Japan}
\affiliation{Institute for Excellence in Higher Education, Tohoku University, 41 Kawauchi, Aoba-ku, Sendai, Miyagi 980-8576, Japan}

\author{K. Ohgushi}
\affiliation{Department of Physics, Graduate School of Science, Tohoku University, 6-3 Aramaki-Aoba, Aoba-ku, Sendai, Miyagi 980-8578, Japan}
\author{J. Chaloupka}
\affiliation{Department of Condensed Matter Physics, Faculty of Science, Masaryk University, Kotl\'{a}\v{r}sk\'{a}  2, 61137 Brno, Czech Republic}
\author{B. Keimer}
\affiliation{Max-Planck-Institut f\"{u}r Festk\"{o}rperforschung, Heisenbergstra\ss e 1, D-70569 Stuttgart, Germany}
\author{H. Suzuki}
\email[]{hakuto.suzuki@tohoku.ac.jp}
\affiliation{Frontier Research Institute for Interdisciplinary Sciences, Tohoku University, Sendai 980-8578, Japan}
\affiliation{Institute of Multidisciplinary Research for Advanced Materials (IMRAM), Tohoku University, Sendai 980-8577, Japan}
\date{\today}

\begin{abstract}
Ruthenium trihalides Ru$X_3$ ($X$ = Cl, Br, I) provide a tunable platform for Kitaev magnetism in two-dimensional van der Waals materials. Despite their similar crystal structures and zigzag antiferromagnetic order, RuBr$_3$ exhibits a higher N\'eel temperature ($T_N$) than RuCl$_3$, suggesting their distinct proximity to the Kitaev quantum spin liquid phase. Using Ru $L_3$-edge resonant x-ray scattering, we show that, while the long-range zigzag order in RuBr$_3$ disappears at $T_N$, the zigzag correlations that persist well above $T_N$ show a pronounced spectral weight redistribution along the interlayer direction. These results suggest that the enhanced interlayer magnetic interactions driven by the extended Br 4$p$ orbitals stabilize three-dimensional zigzag correlations in RuBr$_3$.
\end{abstract}

\maketitle
The Kitaev honeycomb model \cite{Kitaev.A_etal.Ann.-Phys.2006} is not only a canonical example of an exactly solvable quantum spin model but also provides a rigorous realization of a quantum spin liquid \cite{Savary.L_etal.Rep.-Prog.-Phys.2017}. Its potential application in fault-tolerant quantum computation has driven extensive efforts to realize it in physical systems. The theoretical proposal to utilize spin-orbit Mott insulators as a solid-state realization of the Kitaev model \cite{Jackeli.G_etal.Phys.-Rev.-Lett.2009} has spurred significant research into quantum magnetism in transition metal compounds \cite{Rau.J_etal.Annu.-Rev.-Condens.-Matter-Phys.2016,Hermanns.M_etal.Annu.-Rev.-Condens.-Matter-Phys.2018,Takagi.H_etal.Nat.-Rev.-Phys.2019}. 

A major challenge in realizing the Kitaev spin liquid in quantum materials is the competition between the Kitaev interaction and other magnetic interactions. The Kitaev interaction, a bond-dependent Ising exchange, induces strong magnetic frustration that favors a quantum spin liquid ground state. However, additional interactions in real materials, such as the Heisenberg exchange and off-diagonal exchanges \cite{Rau.J_etal.Phys.-Rev.-Lett.2014}, often compete with the Kitaev term and stabilize magnetic orders at low temperatures. In theory, these competing interactions lead to a variety of magnetic ground states, including zigzag, stripy, and spiral orders \cite{Chaloupka.J_etal.Phys.-Rev.-Lett.2010,Rau.J_etal.Phys.-Rev.-Lett.2014}. The zigzag magnetic order has been widely observed in candidate materials, including Na$_2$IrO$_3$ \cite{Hwan-Chun.S_etal.Nat.-Phys.2015} and $\alpha$-RuCl$_3$ (RuCl$_3$) \cite{Sears.J_etal.Phys.-Rev.-B2015}.

Despite the presence of zigzag magnetic order, growing evidence suggests that RuCl$_3$ hosts fractionalized excitations \cite{Sandilands.L_etal.Phys.-Rev.-Lett.2015,Banerjee.A_etal.Nat.-Mater.2016,Banerjee.A_etal.Science2017,Do.S_etal.Nat.-Phys.2017,Banerjee.A_etal.npj-Quantum-Materials2018,Kasahara.Y_etal.Nature2018,Yokoi.T_etal.Science2021,Matsuda.Y_etal.Rev.-Mod.-Phys.2025}, suggesting its proximity to the Kitaev spin liquid phase. A promising route to the spin liquid phase is to fine-tune the magnetism of RuCl$_3$. In this context, the sibling compounds Ru$X_3$ ($X$ = Br, I) \cite{Ersan.F_etal.J.-Magn.-Magn.-Mater.2019,Nawa.K_etal.J.-Phys.-Soc.-Jpn.2021,Ni.D_etal.Adv.-Mater.2022,Imai.Y_etal.Phys.-Rev.-B2022,Prots.Y_etal.Anorg.-Allg.-Chem.2023,Ma.Z_etal.Commun.-Phys.2024,Shen.B_etal.Phys.-Rev.-B2024} offer a platform for the continuous tuning of magnetic interactions. Previous Ru $L_3$-edge resonant inelastic x-ray scattering (RIXS) studies on polycrystalline Ru$X_3$ have confirmed the formation of $J = 1/2$ pseudospins \cite{Gretarsson.H_etal.Phys.-Rev.-B2024}. However, despite their similar in-plane honeycomb structures,  RuBr$_3$ exhibits markedly higher N\'eel temperature ($T_N$) than RuCl$_3$, raising the question of whether this difference originates from the in-plane interactions or enhanced interlayer coupling.

In this Letter, we report a systematic investigation of zigzag magnetic order and fluctuations in RuBr$_3$ single crystals using Ru $L_3$-edge resonant elastic x-ray scattering (REXS) and RIXS. The REXS data confirm long-range zigzag magnetic order, whose intensity exhibits order-parameter-like behavior and almost vanishes at the $T_N$ of 34 K. Meanwhile, the RIXS results reveal that zigzag correlations persisting well above $T_N$ show three-dimensional spectral weight reconstruction along the interlayer direction, indicating that sizable interlayer pseudospin interactions mediated by the spatially extended Br 4$p$ orbitals stabilize three-dimensional magnetic correlations. Contrary to the common assumption that van der Waals materials are two-dimensional (2D), our findings highlight the crucial role of interlayer coupling in van der Waals magnets with heavy ligand ions.

\begin{figure*}[ht]
  \centering
  \includegraphics[angle = 0, width = 1\textwidth, clip=true]{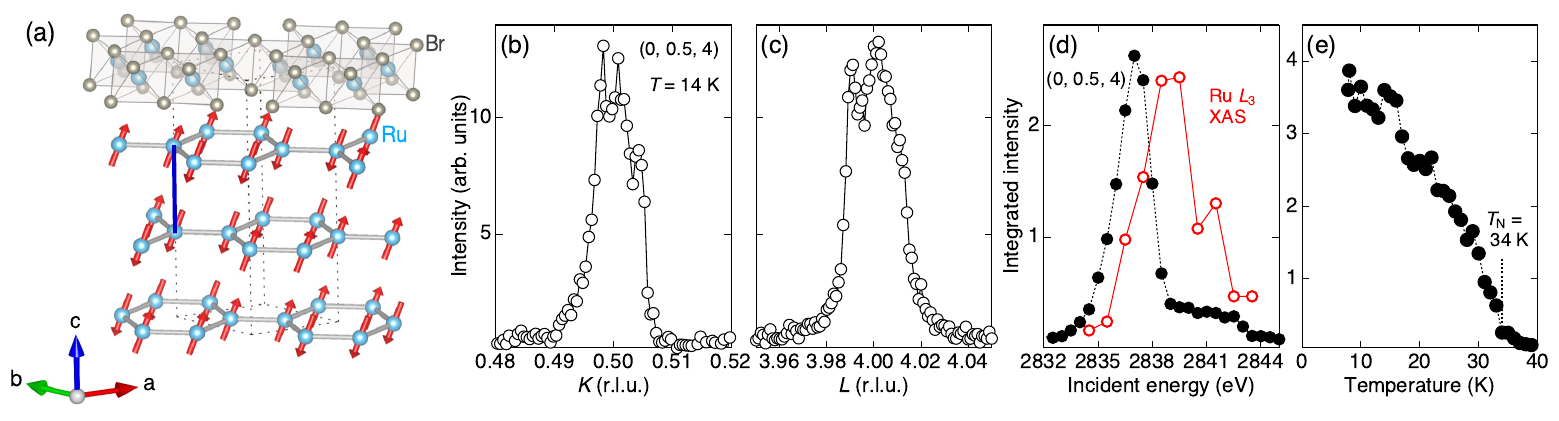}
  \caption{(a) Crystal structure and the zigzag magnetic order of RuBr$_3$, which belongs to the rhombohedral space group $R\bar{3}$. The blue vertical bar indicates the nearest-neighbor interlayer bond. The figure is illustrated using the VESTA software \cite{Momma.K_etal.J.-Appl.-Cryst.2011}. (b), (c) Resonant elastic x-ray scattering (REXS) line scans through the zigzag magnetic Bragg peak $\textbf{\textit{q}} = (0, 0.5, 4)$ at $T = 14$~K taken along the $(0, K, 0)$ [(b)] and $(0, 0, L)$ [(c)] directions.  
  (d) Incident energy dependence of the Bragg diffraction signal (black circles) across the Ru $L_3$-edge x-ray absorption (XAS, red circles).  
  (e) Temperature dependence of the integrated REXS intensity at ${\bm q} = (0, 0.5, 4)$, confirming zigzag magnetic order at the N\'eel temperature ($T_N$) of 34 K \cite{Imai.Y_etal.Phys.-Rev.-B2022}.
  }\label{rexs}
\end{figure*}

The REXS experiment was performed at the beamline P09 of PETRA III at DESY \cite{Strempfer.J_etal.J.-Synchrotron-Radiat.2013} using $\sigma$-polarized light tuned to the Ru $L_3$-edge (2836.5 eV). The polarization of the scattered light was not analyzed. A closed-cycle cryostat was used to cool down the sample, where only one single outer Be dome was used to minimize absorption. The RIXS experiment was conducted using the IRIXS spectrometer \cite{Gretarsson.H_etal.J.-Synchrotron-Rad.2020} at beamline P01 of PETRA III. The measurements were performed with $\pi$-polarized incident photons, while scattered photons of both polarizations were collected at a fixed scattering angle of 90$^\circ$. The setup was similar to a previous study on RuCl$_3$ \cite{Suzuki.H_etal.Nat.-Commun.2021}, but with an improved energy resolution of 75 meV at the Ru $L_3$-edge and cooling capability down to $T = 15$ K. These improvements allowed us to identify the remnant zigzag correlations in RuCl$_3$, which were not resolved in Ref. \cite{Suzuki.H_etal.Nat.-Commun.2021}. The momentum transfer ${\bm q} = (H, K, L)$ is expressed in the reciprocal lattice units (r.l.u.).

We begin by reviewing the crystal and magnetic structures of RuBr$_3$. As shown in Fig.~\ref{rexs} (a), RuBr$_3$ crystallizes in the rhombohedral space group $R\bar{3}$, forming three regular honeycomb layers composed of the edge-shared RuBr$_6$ octahedra that stack in an ABCABC-type sequence. Previous neutron powder diffraction measurements have revealed that the $J = 1/2$ magnetic moments lie within the $ac$ plane at an angle of $\pm$64(12)$^\circ$ from the $a$-axis, but the sign remained undetermined \cite{Imai.Y_etal.Phys.-Rev.-B2022}. As we shall show below, the present RIXS measurements on single crystals provide evidence for ferromagnetic Kitaev interactions, which is consistent with the moment direction of 64(12)$^\circ$ \cite{Chaloupka.J_etal.Phys.-Rev.-B2016}.

We first study the detailed properties of the zigzag order in RuBr$_3$ single crystals by Ru $L_3$-edge REXS measurement. Figures ~\ref{rexs}(b) and (c) present REXS scans along the ${\bm q}=(0, K, 0)$ and $(0, 0, L)$ directions around the zigzag magnetic Bragg peak at ${\bm q} = (0, 0.5, 4)$. The measurement was performed with an incident energy of 2836.5~eV, at $T=14$ K, well below the $T_N$ of 34 K. The data reveal a clear peak, firmly confirming the formation of the long-range zigzag magnetic order. The fine structures indicate the presence of crystalline domains, which is expected in single crystals grown by high-pressure synthesis ~\cite{Imai.Y_etal.Phys.-Rev.-B2022,Zhang.B_etal.Chin.-Phys.-Lett.2025}. Yet, the well-defined peak along the $L$ direction [Fig. \ref{rexs}(c)] indicates strong three-dimensional magnetic correlations, which is contrasted to the broad $L$ dependence in the magnetic diffraction peak in the Ru $L_3$ REXS data of RuCl$_3$ \cite{Sears.J_etal.Nat.-Phys.2020}.

To further establish the magnetic origin of this peak, we examined its resonance behavior and temperature dependence. Figure \ref{rexs}(d) shows the incident energy dependence of the peak intensity across the Ru $L_3$-edge x-ray absorption edge. The data were collected by fixing the scattering angles for ${\bm q} = (0, 0.5, 4)$ and scanning the incident energy. The peak intensity maximum occurs below the peak of the x-ray absorption spectrum (XAS). As the XAS maximum is located at the transition to the Ru $e_g$ orbitals, this incident energy dependence confirms that the zigzag order is associated with the Ru $t_{2g}$ electrons. We also show the temperature dependence of integrated Bragg peak intensity in Fig. \ref{rexs}(e). The intensity exhibits order-parameter-like behavior vanishing near $T_N=34$ K, confirming the disappearance of the long-range zigzag order. Yet, weak residual intensity above $T_N$ is also identified, suggesting the short-ranged zigzag magnetic correlations remaining above $T_N$, which we further investigate using RIXS.

\begin{figure*}[ht]
  \centering
  \includegraphics[angle = 0, width = 1\textwidth, clip=true]{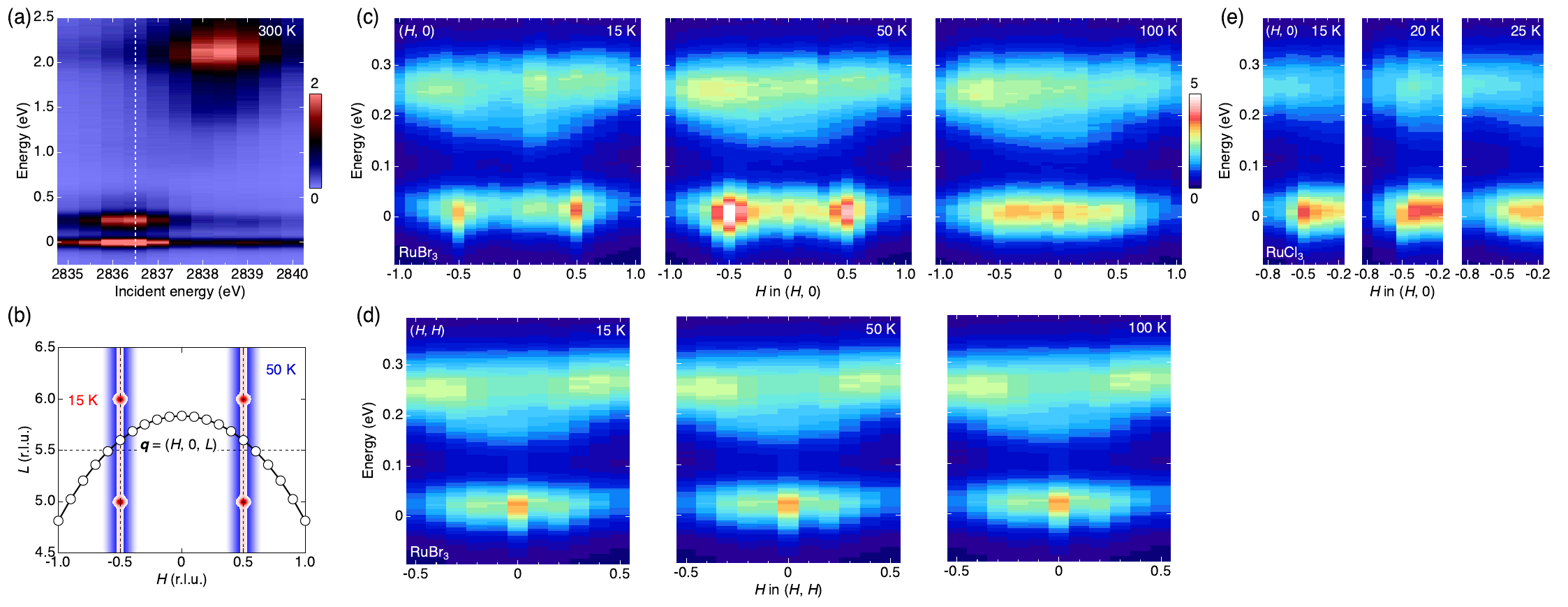}
  \caption{\label{rixs}(a) Colormap of the incident energy dependence of the resonant inelastic x-ray scattering (RIXS) spectra of RuBr$_3$ around the Ru $L_3$ absorption edge. The white dashed line indicates the $t_{2g}$ resonance at the incident energy of 2836.5 eV employed for the momentum-dependent measurements. (b) Schematic illustration of the three-dimensional zigzag correlations in RuBr$_3$ on the ${\bm q}=(H, 0, L)$ plane at 15 K (red) and 50 K (blue). The black empty circles indicate the momentum trajectory for the RIXS measurement. The dashed lines depict the Brillouin zone boundary. (c), (d) Colormap of Ru $L_3$-edge RIXS spectra along the ${\bm q}_{\parallel}=(H, 0)$ and $(H, H)$ directions, collected at $T=$ 15 K, 50 K, and 100 K. (e) Colormap of RIXS spectra of RuCl$_3$ collected at $T=$ 15 K, 20 K, and 25 K.   }\label{map}
\end{figure*}

To resolve the evolution of the zigzag correlations across $T_N$ in RuBr$_3$, we turn to the energy-resolved RIXS technique. In Fig. \ref{rixs}(a) we show the incident energy dependence of RIXS spectra around the Ru $L_3$ absorption edge. The data were collected at $T = 300$ K and at the in-plane wavevector ${\bm q}_{\parallel}=(0, 0)$. The colormap reveals a quasi-elastic intensity around $\omega=0$ eV, a broader feature centered around 250 meV, and a higher-energy excitation emerging around 2 eV. Based on their resonance profiles, these three features can be assigned to magnetic fluctuations within the $J=1/2$ ground states, spin-orbit transitions to the $J=3/2$ states, and crystal field transitions from the $t_{2g}$ to the $e_g$ orbitals \cite{Suzuki.H_etal.Nat.-Commun.2021,Gretarsson.H_etal.Phys.-Rev.-B2024}. The quasi-elastic intensity is most enhanced at the incident energy of 2836.5 eV (white dashed vertical line), which is employed to study the ${\bm q}$ dependence of magnetic correlations. 

To clarify the three-dimensional nature of the zigzag correlations in RuBr$_3$, we performed ${\bm q}$-dependent RIXS measurements.  As the scattering angle of the IRIXS spectrometer is fixed at 90$^\circ$, the measurement path forms a curved trajectory in the reciprocal space. Figure \ref{map}(b) shows the measurement path for the ${\bm q}=(H, 0, L)$ scan (circles), overlaid on the schematic of magnetic fluctuations at 15 K and 50 K. In this trajectory, the in-plane zigzag wavevectors ${\bm q_{\parallel}}=(\pm 0.5, 0)$ correspond to $L=5.6$, deviating from the zigzag magnetic Bragg peaks where $L$ is an integer (red spots in 15 K). By tracking how the spectral weight of magnetic fluctuations is redistributed in the momentum space across $T_N$, we gain insight into the strength of interlayer interactions.

In Fig.~\ref{rixs}(c), we show colormaps of the low-energy RIXS spectra along the ${\bm q}_{\parallel} = (H, 0)$ direction, collected at 15 K, 50 K, and 100 K. At 15 K, well below the $T_N = 34$ K, the quasi-elastic intensity exhibits local maxima near ${\bm q}_{\parallel} = (\pm 0.5, 0)$. Upon warming above $T_N$, the quasi-elastic intensity first increases at 50 K and then becomes weaker at 100 K. This nonmonotonic temperature dependence can be understood as the three-dimensional redistribution of magnetic spectral weight in the ${\bm q}$ space. Below $T_N$, the magnetic scattering is concentrated at magnetic Bragg positions with integer $L$ [see the 15 K profile in Fig.~\ref{map}(b)]. When the long-range zigzag order disappears at $T_N$, these zigzag peaks evolve into continuous rods along the $L$ direction [50 K profile in Fig.~\ref{map}(b)], leading to enhanced spectral weight at non-integer $L$ values and hence increased RIXS intensity at 50 K. At 100 K, the spectral weight develops a broad maximum around ${\bm q}_{\parallel} = (0,0)$, indicating the presence of dominant ferromagnetic interactions.

We also note that the spin-orbit $J = 3/2$ excitations around 0.25 eV remain nearly unchanged across the three temperatures, indicating that the local crystal-field environment of the Ru$^{3+}$ ions is essentially temperature independent up to 100 K. The weak energy minimum near ${\bm q}_{\parallel} = (0,0)$ below the main nondispersive $J = 3/2$ peak is attributed to the formation of exciton bound states \cite{Gretarsson.H_etal.Phys.-Rev.-Lett.2013,Lebert.B_etal.Phys.-Rev.-B2023}.

Figure~\ref{rixs}(d) shows colormaps of the RIXS spectra along the ${\bm q}_{\parallel} = (H, H)$ direction at the three temperatures. In contrast to the $(H, 0)$ direction, the intensity profile along $(H, H)$ displays little temperature dependence, indicating that the reconstruction of spectral weight associated with the zigzag correlations is confined to the vicinity of ${\bm q}_{\parallel} = (\pm 0.5, 0)$.

\begin{figure}[ht]
  \centering
  \includegraphics[angle = 0, width = 0.48\textwidth, clip=true]{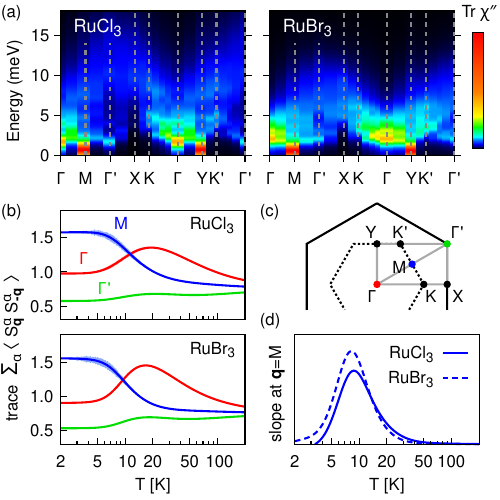}
  \caption{(a) Theoretical dynamical susceptibility $\mathrm{Tr}\chi^{\prime\prime}({\bm q}, \omega)$ of the in-plane Hamiltonian $\mathcal{H}_{\mathrm{2D}}$ of RuCl$_3$ and RuBr$_3$. (b) Temperature dependence of the equal-time spin correlation function $\sum_\alpha \left<S_{\bm q}^\alpha S_{-\bm q}^\alpha\right>$. (c) High-symmetry ${\bm q}$ points in the 2D Brillouin zone and the path used in (a) (gray lines). (d) The temperature derivative of the correlation functions at the M point.}\label{ED}
\end{figure}

To highlight the distinct dimensionality in RuBr$_3$ and RuCl$_3$, we present RIXS spectra of RuCl$_3$ at $T=15$ K, 20 K, and 25 K in Fig. \ref{rixs}(e). In RuCl$_3$, the long-range zigzag magnetic order disappears at $T_N=7$ K \cite{Sears.J_etal.Phys.-Rev.-B2015}. Nevertheless, short-range magnetic correlations persist at $T=15$ K, evidenced by a local intensity maximum around ${\bm q}_{\parallel}=(-0.5, 0)$. As the temperature increases, these zigzag correlations are readily destabilized, as indicated by the rapid suppression of RIXS intensity. This reflects the phase competition between the zigzag and ferromagnetic states in RuCl$_3$ \cite{Suzuki.H_etal.Nat.-Commun.2021}. Concurrently, the spectral weight shifts toward the 2D Brillouin zone center, ${\bm q}_{\parallel}=(0, 0)$, as the magnetic dynamics of the paramagnetic state are determined by the dominant ferromagnetic interactions. 

A comparison of the RIXS results for RuCl$_3$ and RuBr$_3$ reveals a clear contrast in the stability and dimensionality of their zigzag magnetic correlations. The distinct behavior points to enhanced interlayer magnetic interactions in RuBr$_3$, which stabilize robust three-dimensional zigzag magnetic correlations. At the same time, the similar ${\bm q}$ dependence of the high-temperature RIXS intensity in both compounds indicates the similarity of their in-plane magnetic Hamiltonians.

To qualitatively validate this picture, we have derived the in-plane magnetic exchange interactions in RuBr$_3$ based on the density functional theory (DFT) calculations and second-order perturbation expansions from the strong correlation limit \cite{SM}. 
We consider the two-dimensional Hamiltonian on the honeycomb lattice:
\begin{align}
\mathcal{H}_{\mathrm{2D}}&=\sum_{\langle i,j\rangle\in \gamma} \mathcal{H}_{ij}^{(\gamma)} + \sum_{\langle\langle\langle i,j\rangle\rangle\rangle} J_3 {\bm S}_i \cdot {\bm S}_j,\notag 
\end{align}
where the nearest-neighbor (NN) interaction $\mathcal{H}_{ij}^{(\gamma)}$ is expressed as (for $\gamma=z$) \cite{Rau.J_etal.Phys.-Rev.-Lett.2014,Katukuri.V_etal.New-J.-Phys.2014,Rousochatzakis.I_etal.Rep.-Prog.-Phys.2024}
\begin{align}
    \mathcal{H}_{ij}^{(z)}=\  &K S_i^{z}S_j^z
+J  {\bm S}_i \cdot  {\bm S}_j
+ \Gamma (S_i^xS_j^y +
S_i^yS_j^x )  \notag \\
+& \Gamma'(S_i^xS_j^z +S_i^zS_j^x+S_i^yS_j^z + S_i^zS_j^y).
\label{KH}
\end{align}
The third NN Heisenberg interaction $J_3>0$ stabilizes the zigzag order \cite{Rusnacko.J_etal.Phys.Rev.B2019}. 
The derived exchange interactions read $(K, J, \Gamma, \Gamma^\prime, J_3)=(-1.69, -2.47, 2.01, -0.508, 0.126)$ meV, which are in quantitative agreement with previous DFT-based estimates~\cite{Kaib.D_etal.npj-Quantum-Mater.2022}. 
We also analyze the two-dimensional model of RuCl$_3$, $(-5, -3, 2.5, 0.1, 0.75)$ meV, which was obtained from the ${\bm q}$ dependence of RIXS intensity of the $J=1/2$ excitations \cite{Suzuki.H_etal.Nat.-Commun.2021}.

We compute the dynamical susceptibility of $\mathcal{H}_{\mathrm{2D}}$ by the exact diagonalization of finite-size clusters \cite{SM}. Figure~\ref{ED}(a) presents the trace of the zero-temperature dynamical susceptibility, $\mathrm{Tr}\chi^{\prime\prime}({\bm q}, \omega)$, for the RuCl$_3$ and RuBr$_3$ models. The corresponding 2D ${\bm q}$ paths are illustrated in Fig.~\ref{ED}(c). The two models exhibit nearly identical dynamical behavior. The quasielastic spectral weight near $\omega=0$ is maximal at the zigzag wavevectors (M and Y), consistent with the zigzag ground states in RuCl$_3$ and RuBr$_3$. Moreover, both models display ferromagnetic fluctuations around the zone center ($\Gamma$), in line with the broad quasielastic fluctuations around the $\Gamma$ point in the RIXS data [Fig.~\ref{rixs}(d) and (e)].

To reveal the competition between the magnetic phases, we show in Fig.~\ref{ED}(b) the temperature dependence of the equal-time spin correlation function $\sum_\alpha \langle S_{\bm q}^\alpha S_{-\bm q}^\alpha\rangle$ for the two models at three representative momenta (M, $\Gamma$, and $\Gamma^\prime$). At low temperatures, the correlation is maximal at the M point, representing the zigzag magnetic order in RuCl$_3$ and RuBr$_3$. As the temperature increases, it begins to decrease around 5~K, signaling the melting of the long-range zigzag order. Meanwhile, the correlation functions at $\Gamma$ and $\Gamma^\prime$ increase with temperature up to approximately 18~K, reflecting the dominant role of ferromagnetic interactions in determining the magnetic dynamics in the paramagnetic state. Notably, the temperature dependence of the correlation functions is also nearly identical in these two models, reflecting their similar dynamical properties at $T=0$ [Fig.~\ref{ED}(a)]. 

The decrease in zigzag magnetic correlation enables the estimation of $T_N$ for these two-dimensional models. Fig.~\ref{ED}(d) shows the negative temperature derivative of the spin correlation function at the M point. The solid curve for RuCl$_3$ shows a peak at 8.8 K, in good agreement with the experimental $T_N$ of 7 K \cite{Sears.J_etal.Phys.-Rev.-B2015,Kim.S_etal.Phys.-Rev.-B2024,Namba.R_etal.Phys.-Rev.-Mater.2024}. The dashed curve for RuBr$_3$ also shows a peak at 8.2 K. However, this temperature is considerably lower than the experimental $T_N$ of 34 K. This discrepancy suggests that the relative stability of the zigzag order in RuBr$_3$ is not primarily ascribed to the property of the in-plane magnetic Hamiltonian. 

The nearly identical properties of the in-plane Hamiltonians suggest that the interlayer magnetic interactions play a key role in stabilizing the zigzag order in RuBr$_3$. 
The larger spatial extension of the Br 4$p$ orbitals than the Cl 3$p$ orbitals \cite{Choi.Y_etal.Phys.-Rev.-B2022,Gretarsson.H_etal.Phys.-Rev.-B2024} enhances the interlayer hopping integrals between the Br atoms. As a result, the electron hopping between the Ru ions mediated by the Br atoms is enhanced, leading to larger magnetic exchange interactions. 
In the $R\bar{3}$ structure of RuBr$_3$ \cite{Imai.Y_etal.Phys.-Rev.-B2022} and high-quality RuCl$_3$ crystals \cite{Kim.S_etal.Phys.-Rev.-B2024,Namba.R_etal.Phys.-Rev.-Mater.2024}, each spin has one nearest-neighbor interlayer spin [see the nearest-neighbor interlayer bond in Fig. \ref{rexs}(a)]. The interaction on this bond becomes the XXZ-type in the crystallographic $abc$ coordinates, which prefers antiferromagnetic alignment if the out-of-plane spin component is large \cite{Cen.J_etal.Phys.-Rev.-B2025}. The antiferromagnetic interlayer coupling in RuBr$_3$ with the moment angle of 64$^\circ$ appears consistent with this scenario. Indeed, this large moment angle would be unrealistic within the two-dimensional model with the expected hierarchy of interactions \cite{Chaloupka.J_etal.Phys.-Rev.-B2016}, while the interlayer interactions could enhance the out-of-plane component. On the other hand, in RuCl$_3$ with a smaller moment angle of approximately 35$^\circ$ \cite{Cao.H_etal.Phys.-Rev.-B2016}, the out-of-plane component is significantly reduced. Consequently, the relatively weak interlayer interactions are highly frustrated, which could explain the successive magnetic phase transitions driven by an in-plane magnetic field \cite{Janssen.L_etal.Phys.-Rev.-B2020,Cen.J_etal.Phys.-Rev.-B2025}.



In conclusion, we have presented a comprehensive investigation of the zigzag magnetic order and fluctuations in the Kitaev magnet RuBr$_3$ using Ru $L_3$-edge REXS and RIXS. The REXS data firmly establish the formation of long-range zigzag magnetic order. Furthermore, our RIXS results reveal the three-dimensional spectral weight redistribution above $T_N$. These findings indicate that enhanced interlayer pseudospin interactions, mediated by the spatially extended Br 4$p$ orbitals, are responsible for stabilizing the three-dimensional magnetic order. More broadly, our study highlights the crucial role of interlayer coupling in van der Waals magnets with heavy ligand ions, challenging the two-dimensional view of these materials.

We thank K. Nawa and H. S. Kim for enlightening discussions. This work was supported by Grants-in-Aid for Scientific Research from JSPS (KAKENHI) (numbers JP22K13994, JP22H00102,  JP22H01175, JP22K18680, JP23K22446, JP24H01602, JP25K00014, JP25K00955, JP25H01246,  JP25H01247, JP25K07217, JP25K23345), JST CREST (Grant No. JP19198318), the European Research Council under Advanced Grant No. 101141844 (SpecTera), and by the project QM4ST, Grant No.~CZ.02.01.01/00/22\_008/0004572. We acknowledge DESY, a member of the Helmholtz Association HGF, for the provision of experimental facilities. The experiments were carried out at the beamlines P01 and P09 of PETRA III at DESY. Sample growth at ISSP was carried out under the Visiting Researcher's Program (No. 202112-MCBXG-0024 and No. 202205-MCBXG-0072). Computational resources for the ED were provided by the e-INFRA CZ project (ID:90254).

\section*{Data Availability}
The data that support the findings of this article are openly available  \cite{Zenodo}.
\bibliography{RuBr3RIXS}

\end{document}


\title{Supplemental Material for\\ Three-dimensional zigzag correlations in the van der Waals Kitaev magnet RuBr$_3$}

\author{H. Gretarsson}
\affiliation{Deutsches Elektronen-Synchrotron DESY, Notkestrstra\ss e 85, D-22607 Hamburg, Germany}
\author{R. Iwazaki}
\affiliation{Department of Physics, Graduate School of Science, Tohoku University, 6-3 Aramaki-Aoba, Aoba-ku, Sendai, Miyagi 980-8578, Japan}
\author{F. Sato}
\affiliation{Department of Physics, Graduate School of Science, Tohoku University, 6-3 Aramaki-Aoba, Aoba-ku, Sendai, Miyagi 980-8578, Japan}
\author{H. Gotou}
\affiliation{Institute for Solid State Physics (ISSP), University of Tokyo, Kashiwa, Chiba 277-8581, Japan}
\author{S. Francoual}
\affiliation{Deutsches Elektronen-Synchrotron DESY, Notkestrstra\ss e 85, D-22607 Hamburg, Germany}
\author{J. Nasu}
\affiliation{Department of Physics, Graduate School of Science, Tohoku University, 6-3 Aramaki-Aoba, Aoba-ku, Sendai, Miyagi 980-8578, Japan}
\author{Y. Imai}
\affiliation{Department of Physics, Graduate School of Science, Tohoku University, 6-3 Aramaki-Aoba, Aoba-ku, Sendai, Miyagi 980-8578, Japan}
\affiliation{Institute for Excellence in Higher Education, Tohoku University, 41 Kawauchi, Aoba-ku, Sendai, Miyagi 980-8576, Japan}

\author{K. Ohgushi}
\affiliation{Department of Physics, Graduate School of Science, Tohoku University, 6-3 Aramaki-Aoba, Aoba-ku, Sendai, Miyagi 980-8578, Japan}
\author{J. Chaloupka}
\affiliation{Department of Condensed Matter Physics, Faculty of Science, Masaryk University, Kotl\'{a}\v{r}sk\'{a}  2, 61137 Brno, Czech Republic}
\author{B. Keimer}
\affiliation{Max-Planck-Institut f\"{u}r Festk\"{o}rperforschung, Heisenbergstra\ss e 1, D-70569 Stuttgart, Germany}
\author{H. Suzuki}
\affiliation{Frontier Research Institute for Interdisciplinary Sciences, Tohoku University, Sendai 980-8578, Japan}
\affiliation{Institute of Multidisciplinary Research for Advanced Materials (IMRAM), Tohoku University, Sendai 980-8577, Japan}

\maketitle

\section{\NoCaseChange{I. Derivation of pseudospin Hamiltonians from density functional theory calculations}}

Density functional theory calculations were performed using \texttt{Quantum ESPRESSO} \cite{Giannozzi.P_etal.J.-Phys.-Condens.-Matter2009,Giannozzi.P_etal.J.-Phys.-Condens.-Matter2017} with a plane-wave cutoff energy of 100 Ry and a 6$\times$6$\times$6 Monkhorst-Pack ${\bm k}$-point mesh, together with norm-conserving, full relativistic pseudopotentials from \texttt{PseudoDojo} library \cite{van-Setten.M_etal.Comput.-Phys.-Commun.2018}. Wannier functions for the Ru $d$-orbitals were constructed using \texttt{Wannier90} \cite{Pizzi.G_etal.J.-Phys.-Condens.-Matter2020} without enforcing maximal localization. For the evaluation of interaction parameters, we employed \texttt{RESPACK} package \cite{Nakamura.K_etal.Comput.-Phys.-Commun.2021} under the constrained random phase approximation. Since the current version does not support treating spin-orbit couplings, the Wannier functions were reconstructed with scalar relativistic pseudopotentials using the same procedure as above, and the \texttt{wan2respack} interface \cite{Kurita.K_etal.Comput.-Phys.-Commun.2023} was used to convert the Wannier90 outputs. The resulting interaction matrix elements were reduced to spherically symmetric parameters using the Slater-Condon parameters \cite{Sugano.S_etal.1970}, with three quantities taken as independent parameters: the average of the diagonal direct integrals, the exchange integral averaged over the $t_{2g}$ orbitals, and that over the $e_g$ orbitals. Based on these parameters, we constructed a multiorbital Hubbard model and derived an effective low-energy model in the localized limit by performing second-order perturbation theory with respect to the intersite hoppings \cite{Iwazaki.R_etal.Phys.-Rev.-B2021,Iwazaki.R_etal.Phys.-Rev.-B2023}. The experimental structural data for RuCl$_3$ \cite{Park.S_etal.J.-Phys.-Condens.-Matter2024} and RuBr$_3$ \cite{Imai.Y_etal.Phys.-Rev.-B2022} in the $R\bar{3}$ space group were employed for the calculations. 

\section{\NoCaseChange{II. Exact diagonalization calculations of dynamical susceptibility}}

The dynamical spin susceptibility was calculated by the Lanczos exact-diagonalization method \cite{Prelovsek_book_2013} applied to small honeycomb-lattice clusters. Several sizes and shapes of the clusters were employed to improve the momentum resolution by combining the sets of accessible momenta for the individual clusters. Specifically, we have used hexagon-shaped \mbox{24-site} and \mbox{32-site} clusters together with three rectangle-shaped \mbox{32-site} clusters differing in their orientation. The clusters and the accessible momenta can be found in Fig.~8 of Ref.~\cite{Kim.J_etal.Phys.-Rev.-X2020}. The dynamics were evaluated using 500 Lanczos steps, and the final spectra were broadened by Lorentzians with FWHM of $1\:\mathrm{meV}$.

\begin{figure*}[ht]
  \centering
  \includegraphics[angle = 0, width = 1\textwidth, clip=true]{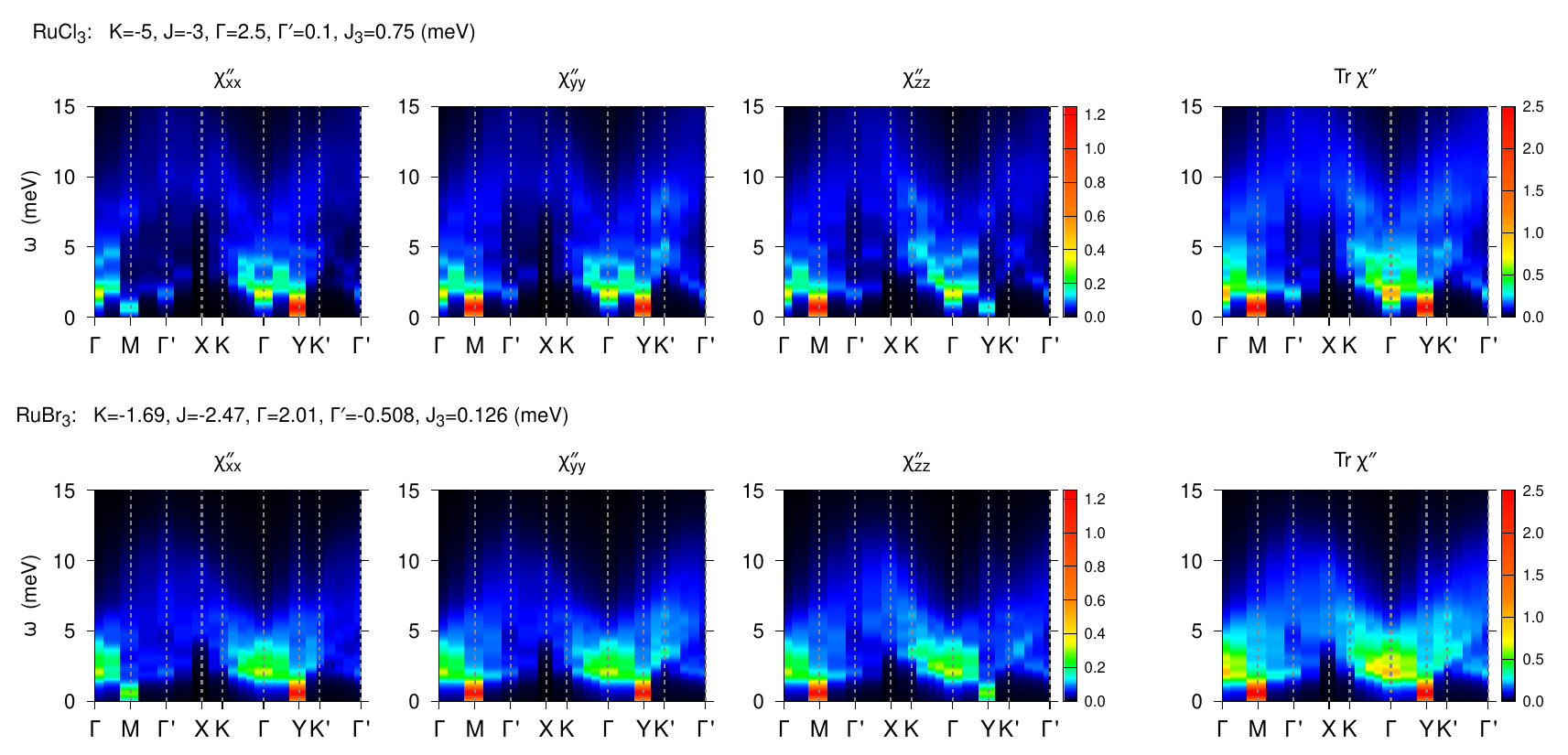}
  \caption{The diagonal components of the dynamical susceptibility $\chi^{\prime\prime}_{\alpha\alpha}({\bm q},\omega)$ ($\alpha=x,y,z$ corresponding to the main axes of the Ru$X_3$ octahedra) at $T=0$ calculated for the two-dimensional RuCl$_3$ model of Ref. \cite{Suzuki.H_etal.Nat.-Commun.2021} (top panels) and the DFT-derived model of RuBr$_3$ (bottom panels). The trace $\mathrm{Tr}\chi^{\prime\prime}({\bm q},\omega)=\sum_\alpha \chi^{\prime\prime}_{\alpha\alpha}({\bm q},\omega)$ is also shown in the right panels.
  }\label{maps}
\end{figure*}

We have computed the dynamical spin susceptibility of the two-dimensional Hamiltonian $\mathcal{H}_{2D}$ for RuCl$_3$ from Ref. \cite{Suzuki.H_etal.Nat.-Commun.2021} and DFT-based one for RuBr$_3$. Figure \ref{maps} shows the diagonal components of the dynamical susceptibility $\chi^{\prime\prime}_{\alpha\alpha}({\bm q},\omega)$ ($\alpha=x,y,z$ corresponding to the main axes of the Ru$X_3$ octahedra) calculated at $T=0$ for these models. The trace $\mathrm{Tr}\chi^{\prime\prime}({\bm q},\omega)=\sum_\alpha \chi^{\prime\prime}_{\alpha\alpha}({\bm q},\omega)$ is also provided in the right panels. Both models contain the pronounced intensity maxima near $\omega=0$ at the zigzag wavevectors (M and Y points). 

The finite-temperature spin correlations presented in Fig.~3(b) and (d) of the main text were obtained for the hexagon-shaped \mbox{24-site} cluster utilizing the finite-temperature Lanczos method in the low-temperature variant \cite{Aichhorn.M_etal.Phys.-Rev.-B2003,Prelovsek_book_2013}. The calculation involved 200 Lanczos steps, and statistical data were collected for 500 random initial state vectors. The statistical error (standard deviation) associated with the method is indicated in Fig.~3(b) by shading. We have also cross-checked these results against the method of thermal pure quantum states \cite{Sugiura.S_etal.Phys.-Rev.-Lett.2012,Sugiura.S_etal.Phys.-Rev.-Lett.2013}.

\bibliography{RuBr3RIXS}